# Modulation of interfacial thermal transport between fumed silica nanoparticles by surface chemical functionalization for advanced thermal insulation

*Takashi Kodama[†*], Nobuhiro Shinohara[‡], Shih-Wei Hung[§], Bin Xu[†], Masanao Obori[†], Donguk Suh[†], Junichiro Shiomi[†*]*

†Department of Mechanical Engineering, The University of Tokyo, Tokyo 113-856, Japan

‡Research Center, AGC Inc., 1150 Uzawa-cho, Kanagawa-ku, Yokohama 221-8756, Japan

§Department of Materials Science and Engineering, City University of Hong Kong, Hong Kong, China

ABSTRACT

Since solid-state heat transport in a highly porous nanocomposite strongly depends on the thermal boundary conductance (TBC) between constituent nanomaterials, further suppression of the TBC is important for improving performance of thermal insulators. Here, targeting a nanocomposite fabricated by stamping fumed silica nanoparticles, we perform a wide variety of surface functionalization on fumed silica nanoparticles by silane coupling method and investigate the impact on the thermal conductivity ($K_m$). The $K_m$ of the silica nanocomposite is approximately 20 and 9 mW/m/K under atmospheric and vacuum condition at the material density of 0.2 g/cm$^3$ without surface functionalization, respectively, and the experimental results indicate that the $K_m$ can be modulated depending on the chemical structure of molecules. The surface modification with a linear alkyl chain of optimal length significantly suppresses $K_m$ by approximately 30%, and the suppression can be further enhanced to approximately 50% with the infrared opacifier. The magnitude of suppression was found to sensitively depend on the length of terminal chain. The magnitude is also related to the number of reactive silanol groups in the chemical structure, where the surface modification with fluorocarbon gives the largest suppression. The surface hydrophobization merits thermal insulation through significant suppression of the TBC, presumably by reducing the water molecules that otherwise would serve as heat conduction channels at the interface. On the other hand, when the chain length is long, the suppression is counteracted by the enhanced phonon transmission through the silane coupling molecules that grows with the chain length. This is supported by the analytical model and present simulation results, leading to predict the optimal chemical structure for better thermal insulation.

INTRODUCTION

Performance improvement of thermal insulation materials used in building and refrigeration units is becoming increasingly important for energy saving[1-6]. The thermal insulator generally takes a highly porous structure composed of low thermal conductivity materials such as silica to suppress the solid heat conduction. The vacuum insulated panel (VIP) is a form of thermal insulator consisting of such porous core media depressurized in a sealing material. As the gas heat transfer is eliminated, VIP can realize low thermal conductivity less than 10 mW/m/K[7-11]. Hence, for applications that require extremely low thermal conductivity, development of advanced materials for VIP with better performance, productivity, and low fabrication cost is important. Since the performance of VIP generally degrades by ageing that gradually lowers the vacuum, nanoparticle-based porous materials are often used for VIP for their relatively robust performance against the gradual loss of vacuum[7,11]. Among the nanoparticle-based thermal insulator, aerogels have been extensively studied, and there are theoretical and experimental reports on the synthesis and physical properties[12-16]. However, since the wet fabrication process for aerogels generally requires supercritical drying that is not suited for production of large-scale materials, simple porous compacts fabricated by dry stamping process are preferable for industrial applications. The porous fumed silica nanocomposite (SNC, the bulk thermal conductivity of silica is approximately 1.38 W/m/K at room temperature)[17-20] is known to be an appropriate material for VIP because the highly porous form shows excellent thermal insulation under vacuum condition (nearly equal to silica aerogels, <10 mW/m/K) and can be simply fabricated by stamping fumed silica nanoparticles (SNP) in dry condition with a small amount of fibers for reinforcement.

The overall thermal conductivity ($K_m$) of SNCs is determined by summation of the effective thermal conductivities by solid ($K_s$), gas ($K_g$), and radiation heat transfer ($K_r$) [21,22]. Here, $K_m$ in VIP can be approximated by $K_m=K_s+K_r$ due to negligible contribution of $K_g$. $K_r$ can be reduced by adding an opacifier with strong optical absorption in the infrared region such as $TiO_2$, SiC, and carbon nanomaterials [20,23-25]. Thus, for further reduction of $K_m$ in vacuum condition, the important challenge is to suppress $K_s$, which is dominantly determined by the interfacial thermal conductance (TBC) between SNPs or between SNP and additives at low material density. An increase in $K_m$ depending on the humidity during the material fabrication process has been reported so far even in vacuum condition, and it may originate from the enhancement of TBC caused by trapped water molecules in the SNC[7]. The hydrophobic surface modification of aerogels has been demonstrated to improve the structural properties, e.g., the porosity and material density, resulting in the enhancement of the insulation performance[23, 26, 27], and it has been reported that material fabrication with hydrophilic or hydrophobic SNPs results in the approximately 30 % variation in $K_m$ (<10 mW/m/K in vacuum when SNPs have hydrophobic properties) due to the difference of adhesion force between silica and reinforced fibers[19]. In contrast, there are several experimental reports showing that the surface functionalization strongly enhances the TBC between solids due to the formation of covalent bonding network at the interface[28-31]. Although these past experimental results support that surface chemical modification is important for the thermal insulation performance, the underlying mechanism has not been clarified yet. Therefore, here we apply a wide variety of surface chemical modifications to SNPs (silanized SNP, SSNP) by a dry silane coupling method[32-35], measure the $K_m$ of SNC fabricated with the SSNPs (silanized SNC, SSNC) by steady-state heat flux method[36], and investigate the detailed impact of surface chemical modification on TBC between SNPs from the measured variation in $K_m$.

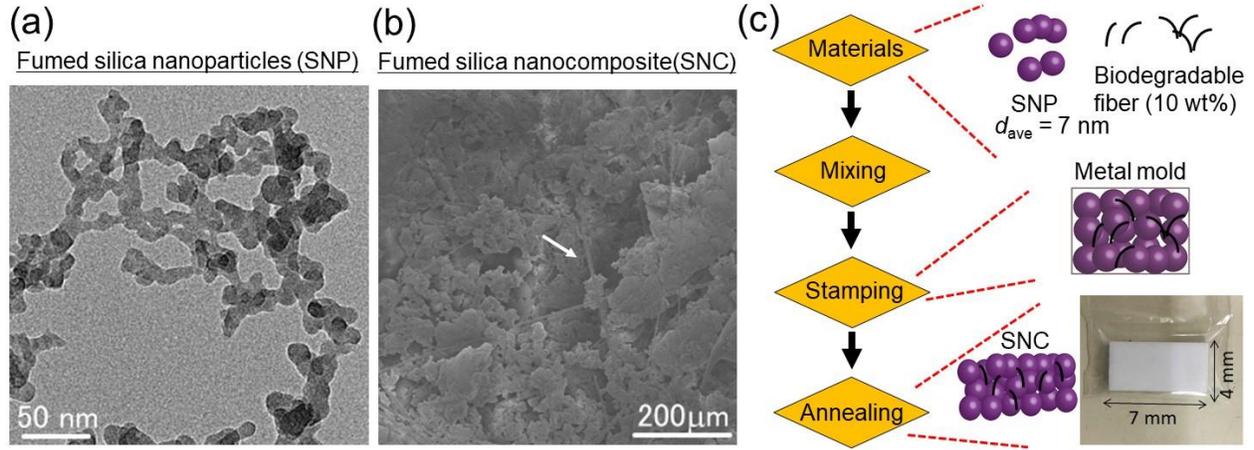

**Figure 1.** Schematics and representative topographic images of SNCs used in present study. **(a)** A transmission electron microscope image of SNPs. **(b)** A scanning electron microscope (SEM) image of SNCs within the volume fraction of 0.072–0.079. The white arrow in the SEM image indicates a biodegradable fiber mixed in SNCs. **(c)** The fabrication procedure of SNCs.

MATERIAL PREPARATION

The measurement samples were prepared with SNPs (or SSNPs) as shown in **Fig.1**. Commercial SNPs (CAB-O-SIL® H-300, Cabot Corporation, average diameter, $d_{ave}$=7 nm) and 10 wt% biodegradable fibers (Superwool PLUS, Shin-Nippon Thermal Ceramics.Corporation) were mixed using a blender. After that, 4×7 mm wide and 10 mm thick SNC was fabricated with a metallic mold by the dry stamping process. The stamping process was performed at room temperature, and small pressure (approximately 1–10 kg/cm$^2$) was applied to the sample during the process to adjust the final sample thickness (10 mm) and target material density. The fabricated materials have a small variation of 0.01 g/cm$^3$ in the material density. The samples were used for the thermal conductivity measurement after being annealed at 200 °C for 2 hours under ambient condition. For suppression of the radiation heat transfer, graphite micro particles (GMP, Nippon Graphite Industries,Co.,Ltd., $d_{ave}$=20 μm) were used as an opacifier. Then, SNPs, 10 wt% biodegradable fibers, and 10 wt% GMPs were mixed by a blender, and the measurement samples were prepared by the same stamping process.

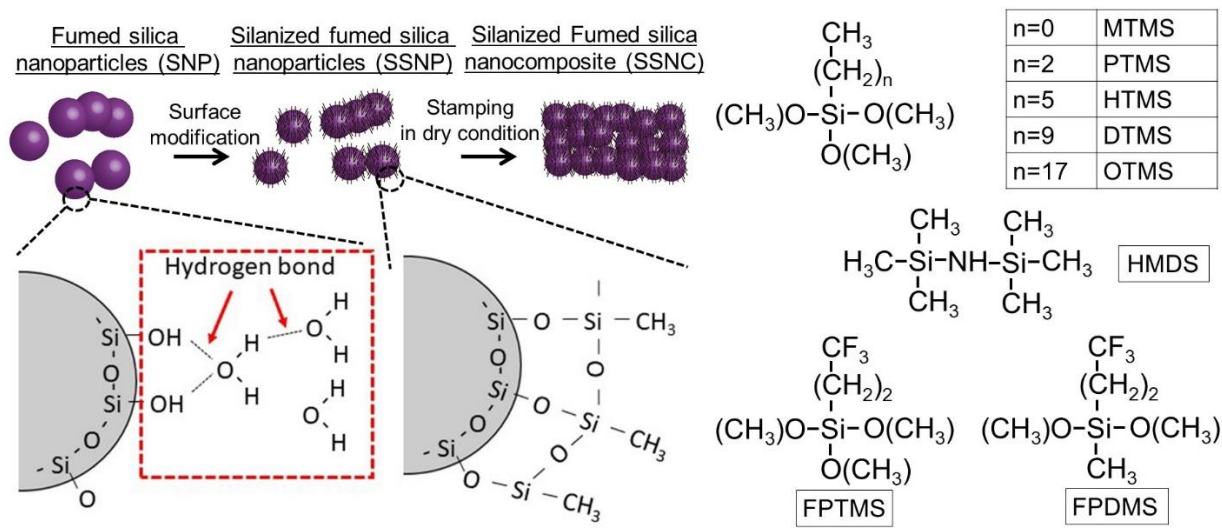

**Figure 2.** Schematics of surface silanization of SNPs, sample fabrication procedure of SSNCs, and structure of chemicals used in present study.

SSNPs were prepared by following silane coupling method in dry condition[34,35]. After cleaning the SNPs by an UV ozone cleaner for ten minutes, approximately 1 g of SNPs and a reservoir including 0.2–0.3 ml silane coupling reagent were sealed with a Teflon container under $N_2$ condition. Then, the container was annealed at 150 °C in a muffle furnace to evaporate the chemicals inside and left for two hours to complete the reaction of evaporated chemicals with hydroxide (-OH) groups terminated on the surface of SNPs. The surface modification process and eight types of chemical structures used in this study are shown in **Fig.2**. For the surface chemical modification in the chain length-dependent experiments, we used five types of silane coupling reagents (Tokyo Chemical Industry Co, Japan) that have different numbers ($n$) of liner hydrocarbon (-$(CH_2)_n CH_3$) groups with three reaction sites (-$SiOCH_3$): methyltrimethoxy silane (MTMS, $n$=0), propyltrimthoxy silane (PTMS, $n$=2), hexyltrimethoxy silane (HTMS, $n$=5), dodecyltrimethoxy silane (DTMS, *n*=9), and octadecyltrimethoxy silane (OTMS, *n*=17). In the experiments, Hexamethyldisilazane (HMDS, Sigma-Aldrich, USA) was also used to react with one -OH group on the silica surface to form -Si(CH$_3$)$_3$ termination. Furthermore, for the experiments with GMPs, we used two types of other silane coupling reagents (Sigma-Aldrich, USA) that have a terminal 3-fluoropropyl group ((-(CH$_2$)$_2$CF$_3$) with two or three reaction sites: 3-fluoropropyl (trimethoxy) silane (FPTMS) and 3-fluoropropyl (dimethoxy) silane (FPDMS). The information on a total of eight chemicals is summarized in **Fig. 2**. We have performed the present dry chemical reaction with flat silica substrates to confirm the surface modification reaction using a contact angle measurement. Moreover, the chemical reaction on SNPs has been also confirmed by FT-IR measurement. Since the coverage ratio of silanes depends on the initial condition of samples, reaction time, and temperature[37], we fixed the reaction condition for the silanization of all samples in this study.

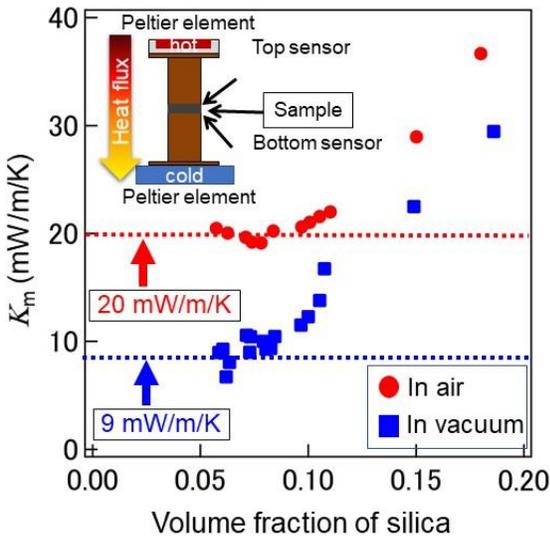

**Figure 3.** The thermal conductivity measurement results of SNCs fabricated in present study as a function of the volume fraction of silica. Dashed lines correspond to the average thermal conductivity within the volume fraction of 0.072–0.079. The inset schematic shows a steady-state heat flux method.

THERMAL CONDUCTIVITY MEASUREMENT

The thermal conductivity of samples was measured by a steady-state heat flux method (see **Fig.3**). In the steady-state heat flux method[22,36], the sample is sandwiched between heat source and heat sink, whose temperature are controlled electrically by Peltier element. The heat flux through the sample and the temperature difference were measured by heat flux and temperature sensors equipped at top and bottom surfaces of the sample after reaching the steady-state, and the thermal conductance was estimated from the measured values (= $q'/\Delta T$). The thermal conductance was evaluated after confirming that the measurement system had sufficiently reached a steady state by controlling the temperature difference between the top and bottom side to be about 10 °C. For the estimation of the thermal conductivity, the actual sample thickness was evaluated experimentally after mounting the samples onto the measurement system. All samples were measured right after the annealing process. All measurements were performed at room temperature in ambient and vacuum conditions (100 Pa).

MOLECULAR DYNAMICS SIMULATION

Non-equilibrium molecular dynamics (NEMD)[38] simulations were carried out to study the interfacial thermal transport across two hydroxylated or alkylsilane-functionalized silica surfaces. Each silica surface is cleaved from α-quartz crystal to form a $5.1 \times 5.3 \times 11.0$ nm$^3$ slab. For silane with different chain lengths, the number of total atoms of simulation system are 47520, 49680, 52560, and 58320 for *n*=2, 5, 9, and 17, repectively. The interactions inside silica surfaces were modeled by using Tersoff potential.[39] The force field parameters, including bond stretching, bond angle bending, and dihedral angle torsion terms for bonded interactions and the van der Waals and Coulomb terms for non-bonded interactions, of hydroxyl groups and alkylsilane chains on surface were obtained from Summers et al.,[40] which is based on the OPLS all-atom model.[41] The geometric combination rule was applied for the parameters of van der Waals interactions between different atoms. The Coulomb interaction was treated using the particle-particle-particle mesh (PPPM) Ewald summation method[42] for slab geometries.[43] The equations of motion were integrated with a time step of 0.5 fs. All the simulations were performed using LAMMPS[44] molecular dynamics package and visualized with the PyMOL[45] software.

The system was initially relaxed in the canonical ensemble (NVT) at 300 K using Nose-Hoover[46,47] thermostat over 1 ns of simulation to ensure the system temperature. After the system reached equilibrium, two 0.5-nm-thick layers at the two ends in the z-direction were fixed to stabilize the system. Then, the NEMD simulations with constant heat exchange algorithm[48] were performed in the microcanonical (NVE) ensemble to calculate the TBC. The two 1.0-nm-thick layers next to the fixed layers at two ends were defined as the heat-source and heat-sink regions, respectively. The constant heat flux, $J$=3.0 GW/m$^2$, was imposed by adding energy to the heat-source regime and subtracting energy from the heat-sink regime. By calculating the temperature difference at the interface, $\Delta T$, the value of TBC, $G$, can be evaluated by $G=J/\Delta T$. A 5 ns production run was performed to obtain the steady temperature profile. A linear regression of the temperature profile of each surface was applied to determine the value of $\Delta T$. The data points near the heat-source and heat-sink regions were excluded from the fitting procedure.

RESULTS AND DISCUSSION

**Figure 3** shows the density dependence of the $K_m$ measured in air and in vacuum condition using a steady-state heat flux method. $K_m$ of a porous structure usually exhibits a minimum value at optimal material density because $K_g$ and $K_r$ increase as the density becomes smaller. As shown in **Fig. 3**, $K_m$ of SNC gives an average value ($K_{0,\text{ave}}$) of approximately 20 mW/m/K in air and 9 mW/m/K in vacuum around the volume fraction ($V_f$) of 0.075 (the material density is 0.2 g/cm³ when $V_f$=0.075). While appropriate mechanical strength is further required for the experiments, the sample has been sufficiently reinforced so that the material can be handled by adding biodegradable fibers at $V_f$=0.075. In present experiments, as there is no significant change in $K_m$ with the moderate variation of fiber concentrations, the fibers play minor role in the thermal conductivity. Further, no specific changes are observed in the topographic image of SSNC after the surface modification of SNPs. Hence, we have confirmed that the $K_s$ is solely dominated by SNPs.

Now that we have confirmed that the $K_s$ is not so sensitive to the $V_f$ around the optimal value of 0.075 and since the purpose of the current work is to identify the dependence of $K_m$ on types of surface functionalizing chemical, we adjust the $V_f$ of the samples discussed hereafter to be in the range of 0.072–0.079. For the same reason, we present the results in terms of the normalized thermal conductivity ($K_m/K_{0,\text{ave}}$) of samples to the reference unmodified SNCs to focus on the variation of surface functionalization effect. At first, as shown in **Fig. 2**, we prepare SSNPs using HMDS and five types of silane coupling reagents that have different length of liner hydrocarbon, and the variation in the $K_m$ of SSNCs are measured experimentally.

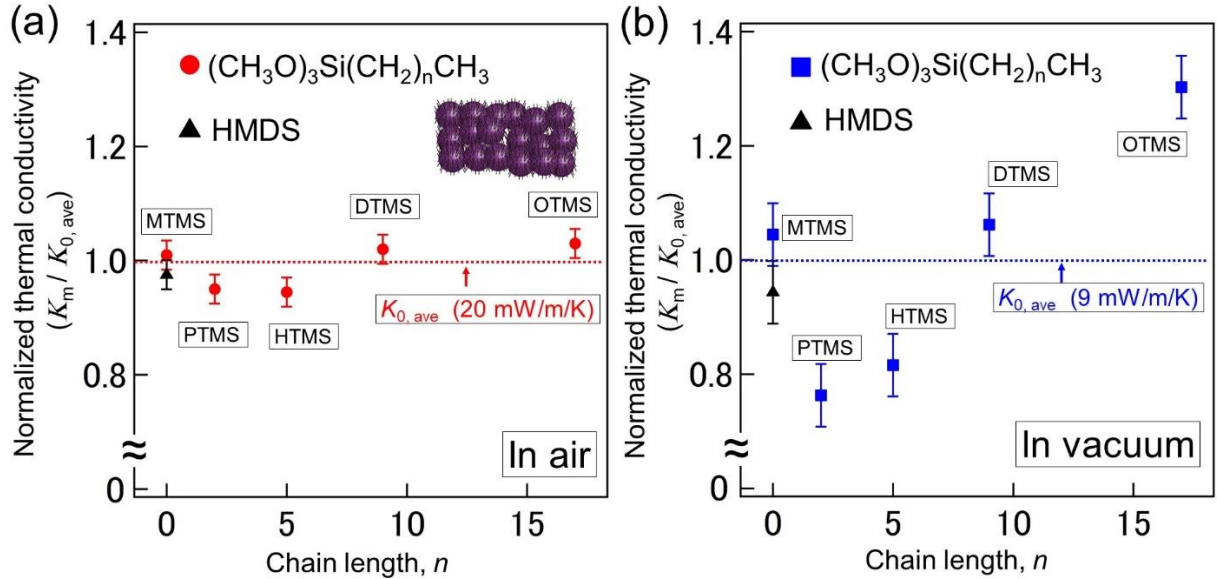

**Figure 4.** Variation in the thermal conductivity ($K_m$) of SSNCs. **(a)** and **(b)** show the normalized thermal conductivity ($K_m / K_{0,ave}$) as a function of the chain length ($n$) of terminated –$(CH_2)_nCH_3$ groups modifying the SNP surface, where $K_{0,\,ave}$ is average thermal conductivity of SNC within the volume fraction of 0.072–0.079. The error bars shown in each data point originate from the uncertainties caused by present steady-state heat flux measurements.

**Figure 4** shows the normalized thermal conductivity ($K_m/K_{0,ave}$) of SSNCs with variation of the chain length of chemicals, $(CH_3O)_3Si(CH_2)_nCH_3$, used for surface modification. As shown in **Fig. 4(a)**, only small variations (<10%) are seen under atmospheric condition. However, **Fig. 4(b)** indicates that the surface modification gives a large variation in $K_m$ under vacuum condition, particularly with shorter silane coupling reagents ($n=2$ and 5). The maximum variation is 20–30% with PTMS ($n=2$), where the corresponding thermal conductivity is 6–7 mW/m/K. Furthermore, $K_m$ increases as the terminal chain of chemicals becomes longer. Although previous reports have mentioned the strong enhancement in TBC through the thiol-based covalent bonding network[28,29], present surface functionalization does not form covalent bond between SNPs. Thus, the observed increase in $K_m$ is caused by different physical phenomena, and these results indicate that the surface modification of SNPs with $(OCH_3)_3Si(CH_2)_nCH_3$ modulates the TBC between SNPs, depending on the alkyl chain length. The surface modification with HMDS shows 5–10% suppression of $K_m$. HMDS is not silane coupling reagent but replaces -OH groups with -Si(CH$_3$)$_3$ on the surface. While the measured variation in $K_m$ is slightly larger than the results with MPTS, which forms -SiCH$_3$ on the surface, the variation in $K_m$ is smaller compared with longer silane coupling reagents. As shown in **Fig.2**, terminal -OH groups on SNP surface can form a stable hydrogen bond with trapped water molecules inside SNCs. Since these modifications make the surface of SNPs hydrophobic, the results suggest that the hydrogen bonding may be the key element that enhances TBC between SNPs. The replaced terminal groups can inhibit the formation of hydrogen bonds and eliminate the trapped water molecules from SSNCs. The -Si(OCH$_3$)$_3$ groups in silane coupling reagents turn to silanol groups (-Si(OH)$_3$) by hydrolysis reaction and the silanol groups are highly reactive with the -OH groups on the surface of SNPs through dehydration reaction. As these chemical reactions create a stable covalent bond network on the surface, the larger number of reaction sites in a silane coupling reagent results in a higher surface coverage that makes the surface more hydrophobic. This gives rise to the larger suppression effect of $K_m$ observed when the SNPs are modified with silane coupling reagents.

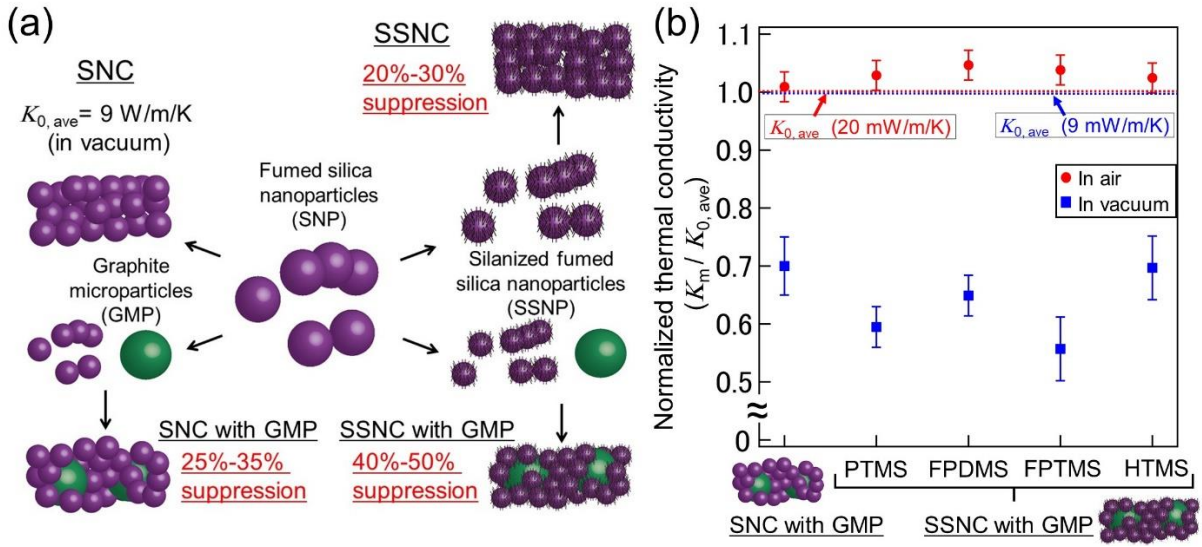

**Figure 5.** Variation in the thermal conductivity ($K_m$) of SSNCs by surface silanization with graphite opacifier. **(a)** Summary of material fabrication procedure and present experimental results. **(b)** The normalized thermal conductivity ($K_m / K_{0,ave}$) in air and vacuum condition with variation of chemicals used for surface silanization.

To further investigate the suppression effect in $K_m$, SNC and SSNC are fabricated with an addition of 10 wt% GMPs as an opacifier to reduce the $K_r$ simultaneously (see **Fig. 5(a)**). **Figure 5(b)** shows the normalized $K_m$ including the infrared opacifier with variation of chemicals. Here we focus on only short silane coupling reagents ($n$=2 and 5) that indicate the greatest suppression effect of the thermal conductivity in **Fig.4**, and four types of chemicals, PTMS ($n$=2), FPDMS ($n$=2), FPTMS ($n$=2), and HTMS ($n$=5), are used for the surface modification. The SNC with GMPs shows 25–35 % suppression in $K_m$ compared with the sample without GMPs in vacuum, whereas no variation is seen in air. The SSNCs with GMPs show larger suppression effect ranging from 40 to 50 % in vacuum with FPTMS, corresponding to thermal conductivity of 4–5 mW/m/K. The experimental results indicate that the surface functionalization is effective strategy for further suppression of $K_m$ with infrared opacifiers. Among the samples, the difference in the measured values between FPTMS and PTMS silanization indicates that a 3-fluoropropyl group is more effective for the suppression of $K_s$. From the comparison of the results between FPDMS and FPTMS, whose only difference is the number of reaction sites in the chemical structure, the silane coupling reagent with three reactive sites gives rise to larger suppression of $K_m$. It is considered that all these findings originate from differences in surface hydrophobicity of SNPs controlled by the chemicals.

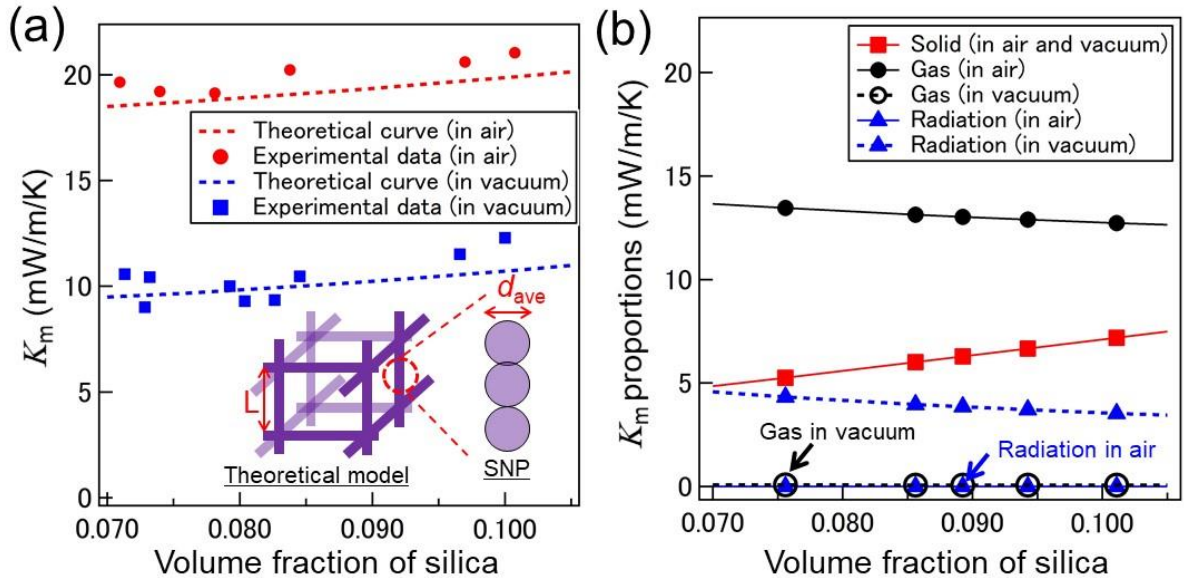

**Figure 6. (a)** Thermal conductivity of SNCs for different volume fractions of silica and the theoretical curve fitted by the theoretical model as reported previously[22]. **(b)** Theoretical contribution of solid, gas, and radiative heat transfer to the overall thermal conductivity in air and vacuum condition as a function of the volume fraction of silica.

We adopt an analytical heat transfer model to gain more microscopic understanding in the experimental results. The details of the model are described in our previous report, where we have validated that the open cell based analytical model can explain well the overall thermal conduction in highly porous nanostructures[22]. In short, the porous media is simply modeled as a cubic lattice structure and a quarter of the cube is defined as a single unit cell in the theoretical model (see **Fig. 6(a)**)[49,50], and the overall thermal conductivity is written by $K_{framework}+K_r$, where $K_{framework}$ is structural contribution of the thermal conductivity including $K_g$. Then, $K_{framework}$ is given by

$$K_{framework} = \frac{4\kappa_{rod}t^2 + K_g(L-2t)^2}{L^2} + \frac{4\kappa_{rod}K_g t^2 + K_g(L-2t)t}{2LK_g t + L\kappa_{rod}(L-2t)} \qquad (1)$$

where $L$, $t$, and $\kappa_{rod}$ are edge length, half of the frame thickness, and thermal conductivity of the SNP rod, respectively. $K_g$ is a proportion of the gas heat conduction given by the commonly used Kaganer model[51,52]:

$$K_g = \frac{\kappa_{g0}}{1+2\beta\Lambda/D} \qquad (2)$$

where $\kappa_{g0}$(=0.026 W/m/K), $\beta$, and $D$ are the thermal conductivity of the gas in free space, a coefficient dependent on the accommodation coefficient and the adiabatic gas coefficient, and the pore size estimated by the volume fraction of the material, respectively. $\Lambda(=k_B T/\sqrt{2}\pi d_g^2 p)$ is the mean free path of gas molecules, where $k_B$, $T$, $d_g$, and $p$ are the Boltzmann constant, the temperature, the gas number density, and the pressure, respectively. The value of mean free path is $1.09\times10^{-7}$ and $1.1\times10^{-4}$ m in air and vacuum condition (=100 Pa), respectively. The proportion of the radiation heat transfer is given by[8,22,53-56]

$$K_r = \frac{16\sigma T^3}{3C(1-\Pi)^f/L} \qquad (3)$$

where $\sigma$ is the Stefan-Boltzmann constant, and $C$ and $f$ are the fitting parameters related to the material property. The $\Pi$ ($=16(t/L)^3 - 12(t/L)^2 + 1$) is a porosity of the material calculated from the density of silica (2.65 g/cm$^3$) and of prepared SNCs that has been determined experimentally. As shown in **Fig. 6(a)**, the quantitative contributions of solid, gas, and radiation heat transfer in $K_m$ can be estimated theoretically by dual fitting (1)−(3) to the density dependent experimental data measured in air and vacuum conditions with seven unknown parameters ($\kappa_{rod}$, $\beta_{air}$, $\beta_{vac}$, $f_{air}$, $f_{vac}$, $C_{air}$, and $C_{vac}$). **Figure 6(b)** indicates the theoretical proportions given by the fitted parameters ($\kappa_{rod}$=0.185, $\beta_{air}$=0.242, $\beta_{vac}$=0.067, $f_{air}$=1.98×10$^{-1}$ $f_{vac}$=1.65×10$^{-1}$ $C_{air}$=2.2×10$^{-1}$, and $C_{vac}$=2.71×10$^{-4}$). This analysis suggests that the $K_m$ is dominated by $K_g$ and the contribution of $K_r$ is much smaller in ambient environment (approximately 5×10$^{-3}$ mW/m/K at $V_f$=0.075) whereas $K_r$ contributes to about 4 mW/m/K in case of vacuum at $V_f$=0.075. In contrast, the contribution of $K_g$ becomes small in vacuum condition (approximately 7×10$^{-2}$ mW/m/K at $V_f$=0.075). The difference of the proportion in atmospheric and vacuum condition reflects the disparate controlling mode for the heat transfer[22]. The analytical results are in good agreement with our present experimental results that the addition of infrared opacifiers to SNCs gives 30–40% reduction in the $K_m$ in only vacuum condition. **Figure 7(a)** shows the theoretical variation of $K_s$ as a function of $\kappa_{rod}$ and $V_f$ when the contribution of $K_g$ is assumed to be neglectable in vacuum ($K_g$ = 0). The $K_s$ (=4 $\kappa_{rod}t^2/L^2$) is nearly proportional to $\kappa_{rod}$. Since the thermal conductivity of solid framework is formed by serially and alternatingly connecting silica with internal thermal conductivity of 1.38 W/m/K and interface with the TBC, the theoretical variation of $\kappa_{rod}$ can be estimated as a function of TBC and contact area ($A_c$) between SNPs ($\kappa_{rod} \approx (1/1.38+A_c/d_{ave}TBC)^{-1}$, see **Fig.7(b)**). The reduction of $\kappa_{rod}$ from 0.185 to 0.135 W/m/K gives rise to the measured 20–30% suppression in $K_m$ at $V_f$=0.075. The reduction of $\kappa_{rod}$ is realized with suppression of TBC to be 50–100 MW/m$^2$K when $A_c$=6–10 nm$^{-2}$.

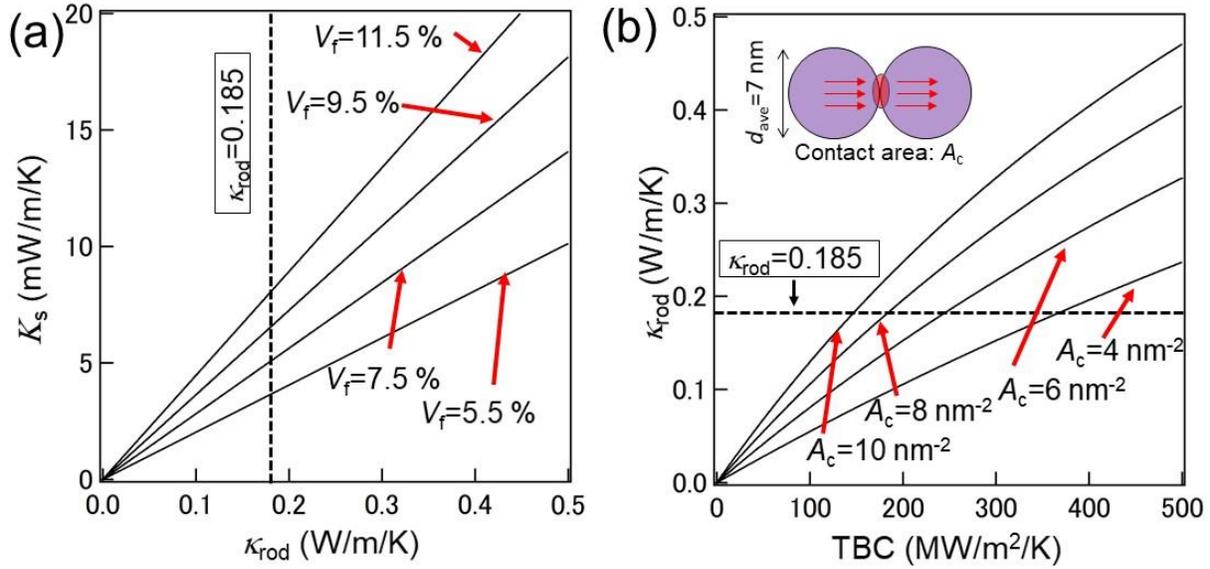

**Figure 7.** (a) Solid component of the overall thermal conductivity as a function of thermal conductivity of SNP frameworks ($\kappa_{rod}$) and volume fraction ($V_f$). (b) $\kappa_{rod}$ as a function of the thermal boundary conductance (TBC) and contact area ($A_c$) between SNPs.

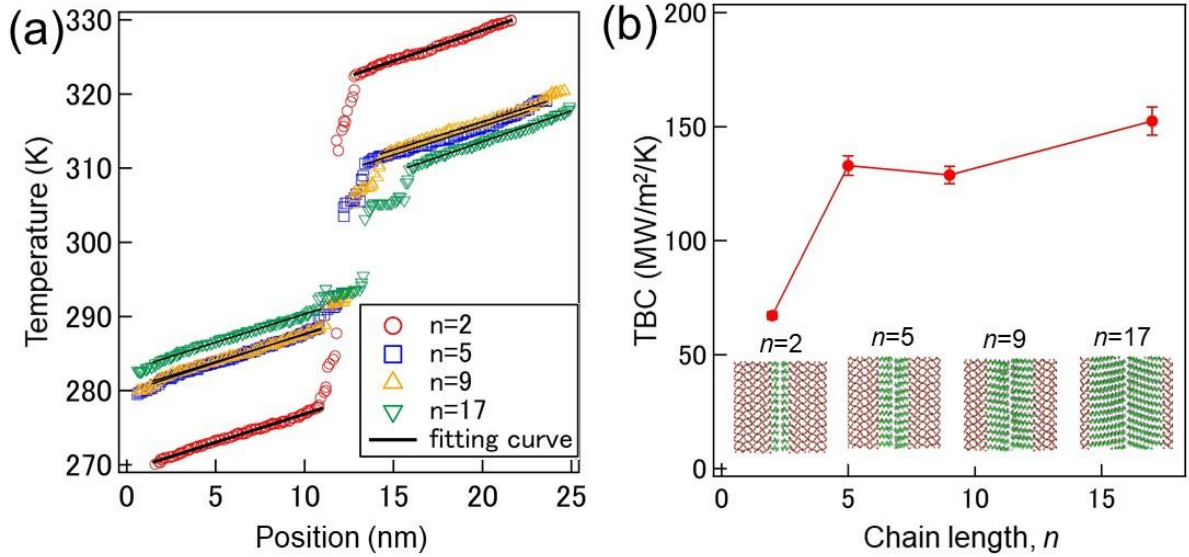

**Figure 8.** (a) Temperature profiles of NEMD simulation for silanized silica with different chain lengths ($n$) of $-(CH_2)_nCH_3$ groups. (b) Resultant thermal boundary conductance of NEMD simulation between silanized silica as a function of $n$.

To further deepen the discussion, a silica-silica interface is modeled with terminal linear hydrocarbon groups with different chain lengths. The variation of TBC through the interface are calculated by NEMD simulations. The temperature profiles with fitted curves of the silanized silica with different chain lengths, $n$=2, 5, 9, and 17, are shown in **Fig. 8(a)**. The slopes of the fitted curves represent the thermal conductivity of silica. The mean value of the resultant thermal conductivity of silica is 3.86±0.17 W/m/K, which is consistent with the previous predicted thermal conductivity of crystalline silica thin films.[57] Clearly, the temperature difference at the interface changes with variation of chain length. The resultant TBC of NEMD simulations are shown in **Fig. 8(b)**, indicating the TBC between SSNPs tends to increase as the terminal – Si(CH$_2$)$_n$CH$_3$ groups implemented on the silica surface become longer. The number of vibration modes of the surface terminal group increases as the molecular weight of functionalized chemicals become larger, and the increase in the overlap of phonon density of state at surface may result in the enhancement of the interfacial phonon transmittance at the interface. This simulation results qualitatively agrees with our experimental results shown in **Fig. 4(b)** that $K_m$ increases as the terminal groups become longer in the range of $n>2$ under vacuum condition. In addition, while the analytical model suggests that the 20–30% suppression in $K_m$ measured for the chain length between n=2 and n=9 result from 50–100 MW/m$^2$K variation in the TBC, the corresponding variation of TBC of approximately 100 MW/m$^2$K calculated from the NEMD simulation is consistent. Therefore, these simulation results can qualitatively explain our present experimental findings. The observed large suppression in $K_m$ is due to the surface hydrophobization that causes a disruption of hydrogen bond network, and the subsequent increase in $K_m$ is due to the enhancement of TBC that originate from an increase in phonon transport between silanized surfaces. Our previous study[58] has also revealed that the TBC at the interface between water and hydrophobic methyl terminated self-assembled monolayers is one order of magnitude smaller than that between water and hydrophilic hydroxyl terminated self-assembled monolayers, which is coincident with the present observation that the thermal insulation is favored by the surface hydrophobization.

CONCLUSIONS

In this study, we target thermal transport properties of porous silica nanocomposite fabricated by dry-stamping process for the application of VIP and demonstrate the systematic surface functionalization to enhance thermal insulation. The surface chemical modification causes a large suppression in $K_m$ (up to 30%) that originates from the elimination of the hydrogen bond network formed by trapped water molecules and has a synergistic effect (up to 55%) with a suppression of radiative heat transfer by addition of infrared opacifiers in vacuum condition. We should note that the surface functionalization is mainly effective under vacuum condition due to neglectable contribution of large $K_g$. The present results predict an optimal chemical structure for better thermal insulation. While surface hydrophobization is effective for the reduction of TBC, the interfacial phonon transport also changes at the same time depending on the chemicals to be modified, which is closely related to the overlapped magnitude of the phonon density of state. Thus, lower molecular weight chemicals that show higher hydrophobicity may be suitable for better suppression of $K_m$. As the reported thermal conductivity of VIP is 4–8 mW/m/K that is nearly equal to the measured thermal conductivity in present experiments[9], and further suppression of the thermal conductivity is still required. However, since the SNC is suitable for production of large-scale materials, the present materials have a strong advantage for the industrial application and the further investigation of the SNC is important for the energy saving society. Fabrication of SSNC consisting of SSNP mixture functionalized with different types of chemicals may be next strategy for the management of interfacial phonon transport for better thermal insulation as the optimal combination is expected to reduce the overlap of phonon density of state at interface.

AUTHOR INFORMATION

Corresponding Author

*E-mail: kodama@photon.t.u-tokyo.ac.jp, shiomi@photon.t.u-tokyo.ac.jp

ACKNOWLEDGMENT

This work was partially support by JST CREST program (JPMJCR16Q5). S.-W.H. was financially supported by JSPS Postdoctoral Fellowship for Overseas Researchers Program (26-04364).

NOTES

The authors declare no competing financial interest.

ABBREVIATIONS

VIP (vacuum insulated panel), TBC(Thermal boundary conductance), SNP(fumed silica nanoparticle), SSNP(silanized fumed silica nanoparticle), SNC(fumed silica nanocomposite), SSNC(silanized fumed silica nanocomposite), GMP(graphite microparticle), MTMS (methyltrimethoxy silane) PTMS(propyltrimthoxy silane), HTMS(hexyltrimethoxy silane), DTMS(dodecyltrimethoxy silane), OTMS(octadecyltrimethoxy silane). FPTMS(3-fluoropropyl (trimethoxy) silane), FPDMS(3-fluoropropyl (dimethoxy) silane). HMDS (Hexamethyldisilazane).

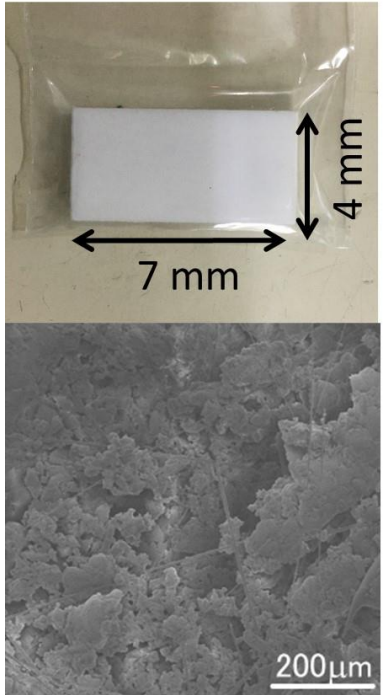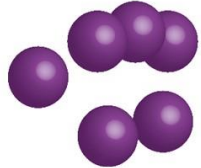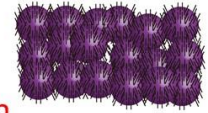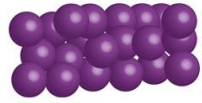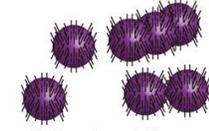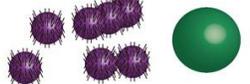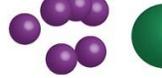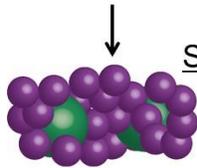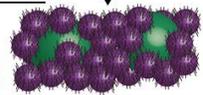

**Table of Contents (TOC) Graphic**

25